\documentstyle[aps,psfig]{revtex}

\def\dxymax{\mbox{$(d\rho_{xy}/dB)^{max}$}}

\def\rxx{\mbox{$\rho_{xx}$}}

\def\sxx{\mbox{$\sigma_{xx}$}}

\def\rxy{\mbox{$\rho_{xy}$}}
\def\T{\mbox{$T$}}
\def\B{\mbox{$B$}}

\def\I{\mbox{$I$}}

\def\Tb{\mbox{$T_{b}$}}
\def\Te{\mbox{$T_{e}$}}

\def\tin{\mbox{$\tau_{in}$}}

\def\up{\mbox{$\uparrow$}}
\def\down{\mbox{$\downarrow$}}
 
\begin{document}
\draft
\preprint{IUCM95-031}

\twocolumn[\hsize\textwidth\columnwidth\hsize\csname
 @twocolumnfalse\endcsname

\title{Phonon Emission from a 2D Electron Gas: Evidence of Transition to
the Hydrodynamic Regime}
 
\author{Edmond Chow, H.P. Wei, S.M. Girvin}
\address{
Department of Physics,
Swain Hall West,
Indiana University,
Bloomington, IN 47405}
 
\author{M. Shayegan}
\address{
Department of Electrical Engineering,
Princeton University,
Princeton, NJ 08544}

\date{\today}
\maketitle
 
\begin{abstract}
Using as a thermometer the temperature dependent magneto-transport of
a two-dimensional electron gas, we find that effective temperature scales
with current as $T_{\rm e} \sim I^a$, where $a=0.4 \pm 2\%$ in
the {\it Shubnikov-de Haas} regime, and $0.53 \pm 2\%$ in both the
{\it integer and fractional} quantum Hall effect.  This implies
the phonon energy emission rate changes from the expected $P\sim T^5$
to  $P\sim T^4$.  We explain this, as well as the dramatic enhancement in 
phonon emission efficiency using a hydrodynamic model. 
\end{abstract}
 
\pacs{PACS Numbers: 73.40.Hm, 73.50.Jt, 73.61.Ey}
\vskip2pc]


The concept of the inelastic scattering time is fundamental to the physics
of disordered quantum systems.  Elastic scattering produces random
interference patterns of the electron waves but does not introduce
decoherence.  In contrast, 
inelastic scattering 
determines the phase coherence time which in turn controls the  
strength of the 
quantum interference observed for example
in the Aharanov-Bohm effect in mesoscopic rings\cite{ABeffect} 
and in Anderson localization in disordered metals.\cite{Localization}  
Inelastic scattering is also fundamental to questions of energy
equilibration and cooling.  In order to see Coulomb blockade
and related effects in mesoscopic devices it is necessary to go to
extremely low temperatures and to isolate the
devices from their environment by means of large series resistances ($R>
h/e^2$). \cite{Webb-and-Lee-book}  It turns out to be difficult to
keep the electrons transiting these thin film resistors from falling out of
equilibrium and heating up\cite{Kuzmin} due to the long inelastic
scattering time at low temperatures. \cite{Roukes}

One difficulty in studying the dynamics of inelastic scattering in ordinary
metallic films is the problem of making and comparing
samples with significantly different characteristics.  We overcome this
problem here by studying high mobility two-dimensional electron gases both
in the low magnetic field Shubnikov-de Haas (SdH) regime and the high field
quantum Hall (IQHE and FQHE) regimes within the same sample.

Previous models of cooling by phonon emission have always assumed that the 
electrons are in plane wave states and invoke {\em momentum conservation}.
Simple phase space considerations\cite{anderson:prl79} then dictate that
the power radiated into phonons is  $P\sim T^{(d+2)}$, where $d$ is
the dimension of space seen by the phonons ($d=3$ in this case).  The
factor $T^d$ arises from the phonon phase space 
and the {\em statically screened}
phonon matrix element. [This applies both to metal films and inversion
layers in piezoelectric media.]
  One additional factor of
$T$ represents the mean energy per phonon and the final factor of $T$
represents the number of electrons which are sufficiently thermally excited
to emit a phonon.  Equating the radiated power to the Joule heating gives
\begin{equation}
T_{\rm e} \sim \left[\rho_{xx} I^2\right]^{1/5} \sim I^{0.4}
\end{equation}
Experimentally we find $T_{\rm e} \sim I^a$ with $a=0.40\pm2\%$ in the SdH
regime.

In the presence of strong disorder or high magnetic field, the reduced mean
free path (and magnetic length) can put the system into a new, hydrodynamic
regime. 
Experimentally we find a new exponent $a=0.53\pm2\%$ 
(corresponding to $P\sim T^4$) and a nearly two order of magnitude
enhancement of the emission rate.  Mittal\cite{mittal} has
discussed the failure of existing theories to properly describe this
regime.      
We present a new hydrodynamic theory of phonon emission in which {\em dynamic
screening} plays a crucial role and which is in quantitative
agreement with these unexpected experimental results.

Current ($I$) or electric field ($E$) dependent transport measurements
have been used extensively to study the electron phonon scattering process
in  2\,DEG's 
 in separate regimes such as
at zero \B\ \cite{wennberg:prb86},  in the SdH regime \cite{sdh},
in between Hall plateaus in the IQHE regime \cite{wei:prb94,koch:sst95},
and at the $\nu = \frac{5}{2}$ FQHE plateau \cite{gammel:prb88}.
The latter indicated a possible dramatic enhancement in the electron-phonon
coupling.  
  However, no experiment has been performed that systematically 
explores different regimes in the same sample  
to explicitly investigate the consequences  of the  QHE,  even though 
it has been suggested that the QHE will affect the electron phonon scattering
\cite{feng:prl93}.
 
The sample used in this work is a GaAs/AlGaAs heterostructure
with an electron density $n=0.65 \times 10^{11}\,\rm cm^{-2}$,
and a mobility $\mu \sim 500,000\rm\,cm^2/Vs$ at 0.1K.
The high mobility was chosen to allow the SdH oscillations to be seen
at low B.  
The donor 
spacer thickness is 1800\AA, and
the Hall bar pattern on the sample has a channel width of 300\,$\mu$m.
The large device is used to avoid the influence of edge effects
\cite{zheng:prl85}, 
which are known to be more pronounced in small devices,\cite{mceuen:prl90}
and to avoid problems with heat being carried out of the sample directly
by electron diffusion.\cite{mittal}    
The sample and a ruthenium oxide resistor, to measure
the temperature (\Tb) at the sample position, 
are mounted in a dilution refrigerator [10,11(c)]
for 45\,mK$<\Tb<$1\,K.
The transport coefficients are measured by 
the standard ac lock-in technique.
It is known that low-frequency 
ac and dc current produce the same results for the
current dependence [10,11(b)].  
 
The inset of\ Fig.1 shows 
\rxx\ as a function of \B\ for
0.06\,T $< B <$ 0.16\,T (which corresponds 
to spin-unresolved Landau level filling factors,
$18 \leq \nu \leq 38$) at 3 different excitation
currents $(I)$ at fixed $\Tb=100$\,mK.
We measure the $\Tb$ 
dependence of these SdH oscillations for
$\rm 100\,mK < \Tb < 750\,mK$
while applying $I=5$\,nA.
By fitting the \T\ dependent \rxx\ amplitudes
to Ando's semi-classical formula \cite{ando:jpsj74}, we 
obtain the electron effective mass $0.068\pm0.001$ 
in units of the free electron mass,
comparable to that obtained from cyclotron resonance 
measurements \cite{chou:prb88}.   

Using the \T\ dependent \rxx\ amplitude as a thermometer,
we find the same relation between \Te\ and $I$ 
for all oscillations down to 0.092\,T in the inset of Fig.1. 
Oscillations at lower \B\ exist only in a very small range of \I.
We choose to plot, in Fig.1, \Te\ vs.\ $I$ for the two oscillations 
pointed to by arrows in the inset of Fig.1, 
because their amplitudes are measurable in a wider range of $I$. 
The closed and open symbols represent 
the data taken from the peak at $B=0.139$\,T 
and the dip at $B=0.133$\,T respectively.
Different symbols represent data taken at 
different \Tb's which are labeled by the side of the data set.
For clarity, the open symbol data set has been offset by scaling 
by a factor of 1/3. 
For each \Tb, there is a current $I_0$, below which \Te\ remains constant,
and above which \Te\ merges into a single power law:
$ T_{\rm e} \sim I^a $
over about one and a half decades in $I$ with $a$ close to 0.4.
The solid line under each data set has a slope of 0.4 
and is drawn for reference.
For comparison, we also draw a dashed line of slope 0.5.  
The experimental value of $a$ is obtained by first
collecting all data points which have $\Te > 2 \Tb$ for each fixed \Tb,
and then performing a linear least-squares fit to the resulting data points.
We obtain $a = 0.4 \pm 2\%$, where 2\% sets the statistical 68\%
confidence limit.

We also deduce the absolute energy relaxation rate ($1/\tin$)
by equating the input Joule heating 
to the cooling rate $P \simeq C T_{\rm e}/5\tin$,   
where $C$ is the specific heat of the 2\,DEG at $B=0$.  
We find $1/\tin \sim 7.3 \times 10^8 T^3\,\rm sec^{-1}K^{-3}$ at $B=0.13$\,T.
Price has calculated   
$1/\tin = 2.92\times 10^8 T^3\,\rm sec^{-1}K^{-3}$ 
for screened 
piezoelectric coupling in the absence of 
disorder.\cite{price:jap82}
This agreement indicates
that statically screened  
piezoelectric electron-phonon coupling 
controls the physics in the SdH regime. 
 
In the same sample, we also study the effect of \I\ 
on \rxy\ in the FQHE regime \cite{note}.        
The inset of\ Fig.2 shows \rxy\ as function of \B\ 
at 3 different \I's with fixed $\Tb=100$\,mK,
between well developed FQHE states at $\nu= \frac{1}{3}$ and $\frac{2}{5}$.
There is no 
signature of higher order fractional states 
for $\frac{1}{3} < \nu < \frac{2}{5}$ in our \T\ range.
We characterize the effect of  

\begin{figure}[t]
\centerline{
\psfig{figure=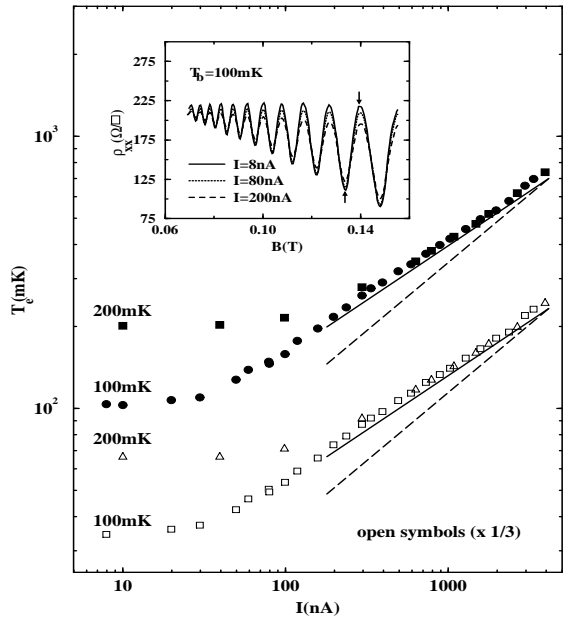,height=3.5in,width=3in,angle=0}}
\caption{
     Electron effective temperature \Te\ vs.\ applied current \I\
     at \Tb= 100 and 200\,mK,   %
     for $B=0.139$\,T (closed symbols)
     and $B=0.133$\,T (open symbols).
     The solid line under each data set
     has slope 0.4, the dashed line 0.5.
     Inset: \rxx\ vs.\ \B\ at $\Tb = 100$\,mK for
     three different $I$'s.
     }
\end{figure}

\noindent
\I\ by assigning a \Te\ at each \I,
where \Te\ is obtained by using \dxymax\ as a thermometer.
Here, \dxymax\ is 
the maximum slope of \rxy\ vs.\ \B\ for $\frac{1}{3} < \nu < \frac{2}{5}$
measured with $I=3$\,nA.
In\ Fig.2, we plot \Te\ vs.\ \I\ 
in this region using open symbols, scaled by a factor of 1/6. 
Different symbols represent 
data taken at different \Tb's or upon 
different cooldowns of the sample. 
We also study the IQHE regime for
Landau level index 
 $N=2\downarrow$ ($5< \nu < 6$) 
and $2\uparrow$ ($4 < \nu < 5$), 
where \up\ and \down\ represent spin direction.
We do not study the IQHE transition for $N=1\downarrow$ and $1\uparrow$
because there appears to have FQHE structures in \rxy.
The result for $N=2\downarrow$ is plotted in\ Fig.2 by closed symbols.
There are no data points for $I>2$\,$\mu$A for the FQHE data
because \dxymax\ is saturated to the classical Hall resistance.
However, it is still clear that in both the IQHE and FQHE regimes
\Te\ behaves the same and there is a single power law, 
$T_{\rm e} \sim I^a$,
over more than one decade 
in \I\ with $a \sim 0.5$.
The solid line of slope 0.5 and the dashed line of slope 0.4
are drawn for reference.
We obtain the best fit
values $a =0.53 \pm 2\%$ 
in the FQHE regime and     
$a =0.54 \pm 2\%$ and $0.53 \pm 4\%$ for 
$N=2 \downarrow$ and $2 \uparrow$ respectively.
This is {\it significantly different} 
from that ($0.40 \pm 2\%$) in the SdH regime.
Similar results have also been obtained in a second sample.
This change in $a$ can be viewed as a change in the cooling rate
from $P \sim T^5$ in the SdH to $P \sim T^4$ in the QHE. 

We note that in our sample the peak values of 
\rxx\ are 212, 210, 335,   
and 4620\,$\Omega/\Box$ for $B=0.139$\,T, $N=2\downarrow$, 
$N=2\uparrow$, and $1/3 < \nu < 2/5$ respectively at $T=100$\,mK.

\begin{figure}[t]
\centerline{
\psfig{figure=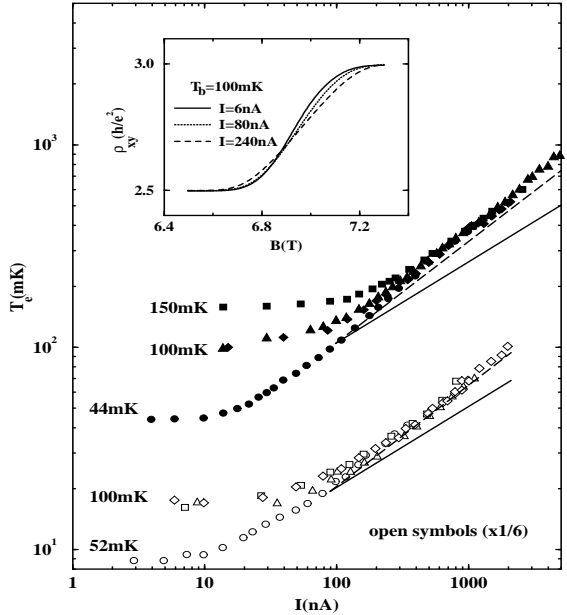,height=3.5in,width=3in,angle=0}}
\caption{
     Electron effective temperature \Te\ vs.\ applied current \I\
     in the IQHE for $5< \nu < 6$ (closed symbols),
     and in the FQHE for $1/3 < \nu < 2/5$ (open symbols).
%
     Each curve is labeled by the bath temperature.
     The dashed reference line under each data set
     has a slope of 0.5, and the solid line 0.4.
     Inset: \rxy\ vs.\ \B\ at $\Tb = 100$\,mK for
     three different $I$'s in the FQHE for $1/3 < \nu < 2/5$.
     }
\end{figure} 

\noindent
The $\rho_{xx}$ and hence the Joule heating at the
fractional transition is 22 times that at the integer transition 
shown in Fig.2, and yet the two curves are nearly identical.  
In addition, the current threshold in Fig.2 for
seeing heating at $T_b =100$\,mK is about 
twice as large as for the SdH case in Fig.1.  This 
indicates that, at the integer transition, cooling is about 4 times as
efficient as in the  SdH regime, and at the fractional transition,
fully 80 times as efficient.    
  One possible explanation for 
this dramatic enhancement and the $P \sim T^4$ power law 
is outlined in the following hydrodynamic model.

The standard Fermi's Golden Rule expression for the rate of phonon 
emission\cite{price:jap82,dassarma} can be rexpressed in the following form
to yield the power radiated by the 2DEG at temperature $T_{\mathrm e}$ into
the lattice at $T_{\mathrm b} = 0$:
\begin{equation}
P = \sum_{\lambda,{\mathbf Q}}
\omega_\lambda({\mathbf Q})
|M_\lambda({\mathbf Q})|^2 n_{\mathrm B}(\beta_{\mathrm e}\hbar
\omega_\lambda({\mathbf Q})) G(q,\omega_\lambda({\mathbf Q})),
\label{eq:thepower}
\end{equation}
\goodbreak
where $G(q,\omega)$ is given by
\begin{equation}
 \frac{4\pi\kappa\epsilon_0 q}{2\pi e^2}\left\{-2
{\mathrm Im}\frac{1}{\epsilon(q,\omega)}\right\}\approx
\frac{2\omega(2\kappa\epsilon_0)^2}{e^2}
{\mathrm Re}\left(\frac{1}{\sigma_{\mathrm xx}}\right),
\end{equation}
$\omega_\lambda({\mathbf Q})$ is the phonon frequency for polarization
$\lambda$, ${\mathbf q}$ is the projection of the wavevector ${\mathbf Q}$
onto the plane of the 2DEG, $M_\lambda({\mathbf Q})$ is the electron-phonon
matrix element, 
$n_{\mathrm B}$ is the Bose-Einstein factor,
$\beta_{\mathrm e}=1/k_{\mathrm B}T_{\mathrm e}$,
$\kappa = 12.9$ is the background semiconductor dielectric
constant and $\epsilon(q,\omega)$ is the 2DEG dielectric function.  The
expression for $G$ in terms of the conductivity
$\sigma_{\mathrm xx}$ is the standard one used
in the theory of surface acoustic wave absorption.\cite{SAW}
 
We assume that at high \B\ and low \T\ the system is 
in the hydrodynamic limit where typical
phonon wave vectors and frequencies are negligible compared to
characteristic scales over which the conductivity varies, so that \sxx\ 
is a real constant. 
This is justified from direct measurement \cite{engel:prl93}
that at the IQHE critical point the peak value of \sxx\ (at $q\sim 0$) is
approximately independent of frequency  
up to {\em at least} 14\,GHz ($\hbar\omega\sim 0.7 K$). 

At low temperatures, piezoelectric coupling
dominates\cite{price:jap82,dassarma} so that $|M_\lambda({\mathbf Q})|^2
\sim 1/Q$. Using the fact that \sxx\ is constant,
simple power counting in Eq.(\ref{eq:thepower}) yields $P\sim T^4$.
The physical interpretation of this is the following.  In the
hydrodynamic limit, momentum conservation is lost because of frequent
collisions of the electrons with the disorder potential.  The rate energy
is absorbed from
 a phonon mode depends only on the square of its electric field and is
independent of wave vector and frequency.  Hence the 2DEG acts like a black
(more precisely a ``gray'') body and emits phonons with the usual $T^4$
spectrum.

In the clean limit, the power counting is different since the conductivity
is non-local:
${\mathrm Re}(1/\sxx) \approx {q}/{q_{\mathrm TF}2\kappa\epsilon_0
v_{\mathrm F}}$, where $q_{\mathrm TF}$ is the Thomas-Fermi wave vector,
$v_{\mathrm F}$ is the Fermi velocity.
This changes the temperature exponent to $P\sim T^5$ and
correctly reproduces the usual Thomas-Fermi static screening
result\cite{price:jap82} for the prefactor.
In the hydrodynamic limit, the static screening approximation fails
(since charge fluctuations are no longer high frequency plasmons at 
$\omega \sim q^{1/2}$ but rather 
relax slowly with $z=1$ dynamics: $\omega = 
i(\sxx/2\kappa\epsilon_0)q$).    
It is not the static compressibility of the electron gas that counts, 
but rather its limited ability to move charge quickly.
The smaller the conductivity is, the poorer the screening is and hence the
larger the emitted power.  Using $\sxx = e^2k_{\mathrm F}\ell_{\mathrm e}/h$ 
gives an elastic mean free path of 
$\ell_{\mathrm e} \approx 1.9 \mu{\mathrm m}$
at $B=0$ and so the system
enters the hydrodynamic regime $q\ell_{\mathrm e}<1$ only at temperatures
below $\sim 12{\mathrm mK}$.  
If however $\sxx \sim e^2/h$, the hydrodynamic regime extends up to about
$1{\mathrm K}$.  

Evaluation of Eq.(\ref{eq:thepower}) yields
\begin{equation}
P = \frac{e^2}{h\sxx}6.75\times10^{-2}\left(\frac{T_{\mathrm e}}{1{\mathrm
K}}\right)^4 {\mathrm Watts/m}^2.
\end{equation}
This expression,  which contains no adjustable parameters,  
 yields the correct \T\ dependence and (using the measured values of
$\rho_{xx}$ and $\rho_{xy}$) lies below
the experimentally measured power by only 17\% (15\%)  
for the IQHE (FQHE) transition.
Considering the simplicity of the model, this level of agreement 
is quite good.  Notice that one of
the advantages of this formulation is that, unlike previous formulations,
 it is not necessary to make any
assumptions about the unknown density of states in the interacting 
2\,DEG.           

Equating the Joule heating and radiated power we have
\begin{equation}
T_{\mathrm e} = 24.9 (\sigma_{xx}\rho_{xx})^{1/4} 
({\mathrm J}/{\mathrm A/m})^{1/2} {\mathrm K},
\end{equation}
which is $\sim 4\%$ 
below the observed values.
This result explains  
why the curves for the fractional and integer transitions are
nearly identical. 
The factor $(\sigma_{xx}\rho_{xx})^{1/4}$ (which is unity 
at $B=0$) is a weak function of $B$ in the sense that it is
nearly equal for the two plateau transitions
 studied.\cite{caveat_smg}    
Thus the increased Joule heating
is precisely compensated by increased electron-phonon coupling
due to reduced screening.   That is, the threshold for heating occurs,
to a good approximation, at fixed current rather than fixed power.
 
We note that
the result $a = 0.53 \pm 2\%$ is close to the prediction
$a = 0.5$ by Polyakov and Shklovskii   
in their theoretical model of \T-scaling \cite{polyakov:prl93}.
They consider the insulating IQHE plateau regime  
where Coulomb variable range hopping dominates the transport.
The electron-phonon interaction is not directly relevant in that study.

The present hydrodynamic heating model for the metallic critical point
assumes that phonon emission is the
rate limiting step and that the electrons are effectively in equilibrium at
some temperature.  An alternative picture can also be developed
in which one recognizes that there is a diverging correlation time
at the critical point of the quantum Hall transition.\cite{sondhi,steve}
Assuming that it is this internal
time scale rather than phonon emission that controls the dynamics, it
can be shown\cite{steve} that $a = z/(1+z) = 0.5$, also
in agreement with the present experiments   
(since for Coulomb interactions, one expects the dynamical exponent $z=1$).
More theoretical and experimental work will be needed to distinguish
these two models.\cite{steve}
One caveat for the quantum critical  picture is that, 
because of the high mobility, the sample does not show critical point power-law
\T-scaling in the linear response transport in the \T\ range studied.   
 
Finally we note that in silicon deformation potential coupling yields $P\sim
T^6$ and hence an exponent $a = 1/3$.  This may well explain the
apparent failure of dynamical scaling in
the experiments of Kravchenko et al.\cite{Furneaux}
on non-linear response in Si inversion layers where $a=1/3$ is observed.

We thank Drs.\ Lloyd Engel, Allan MacDonald, 
S.\ Das Sarma, S.\ Sondhi, M.P.A.\ Fisher,
S.\ Kivelson, T.\ Brandes, and D.C.\ Tsui for stimulating discussions.
The work at Indiana University
is supported by IU Foundation and NSF grants DMR-9311091 and DMR-9416906.
MS at Princeton is supported by NSF DMR-9222418.


\begin{references}
\bibitem{note}   
        We choose \rxy\ as our thermometer in the QHE because
        the evaluation of \dxymax\ only involves one position
        in \B, whereas the width of \rxx\ requires two.
        Both produce the same result \cite{koch:sst95}.
\bibitem{caveat_smg}  On the Hall plateaus this quantity is a strong
function of temperature and frequency and the present analysis would have to be
modified for that case.  %
\end{references}
\end{document}